\documentclass[prl,nofootinbib]{revtex4}
\usepackage{graphicx}
\usepackage{latexsym}
\def\be{\begin{equation}}
\def\ee{\end{equation}}
\def\bea{\begin{eqnarray}}
\def\eea{\end{eqnarray}}

\begin{document}
\title{Probing for cosmological parameters with LAMOST measurement}

\author{Hong Li$^{a,b,c}$}
\author{Jun-Qing Xia$^{a,d}$}
\author{Zuhui Fan$^{c}$}
\author{Xinmin Zhang$^{a,b}$}

\affiliation{${}^a$Institute of High Energy Physics, Chinese Academy
of Science, P.O. Box 918-4, Beijing 100049, P. R. China}

\affiliation{${}^b$Theoretical Physics Center for Science Facilities
(TPCSF), Chinese Academy of Science, P.R.China}

\affiliation{${}^c$Department of Astronomy, School of Physics,
Peking University, Beijing, 100871, P. R. China}

\affiliation{${}^d$Scuola Internazionale Superiore di Studi
Avanzati, Via Beirut 2-4, I-34014 Trieste, Italy}

\begin{abstract}
In this paper we study the sensitivity of the Large Sky Area
Multi-Object Fiber Spectroscopic Telescope (LAMOST) project to the
determination of cosmological parameters, employing the Monte Carlo
Markov Chains (MCMC) method. For comparison, we first analyze the
constraints on cosmological parameters from current observational
data, including  WMAP, SDSS and SN Ia. We then simulate the $3$D
matter power spectrum data expected from LAMOST, together with the
simulated CMB data for PLANCK and the SN Ia from 5-year Supernovae
Legacy Survey (SNLS). With the simulated data, we investigate the
future improvement on cosmological parameter constraints,
emphasizing the role of LAMOST. Our results show the potential of
LAMOST in probing for the cosmological parameters, especially in
constraining the equation-of-state (EoS) of the dark energy and the
neutrino mass.
\end{abstract}

\maketitle

\section{Introduction}

The measurement of the large scale galaxy clustering has been an
important probe in constraining the cosmological models. The large
scale structure (LSS) measurements have made remarkable progresses
by the observational efforts such as 2dFGRS and Sloan Digital Sky
Survey (SDSS), which have provided an accurate measurement of the
galaxy power spectrum and given a robust constraint on cosmological
parameters \cite{Tegmark:2006az,Cole:2005sx}.

The LAMOST \cite{lamost} project is a 4m quasi-meridian reflecting
Schmidt telescope laid down on the ground. It has a 5 degree field
of view, and may accommodate as many as 4000 optical fibers and the
light from 4000 celestial objects will be led into a number of
spectrographs simultaneously. Thus the telescope will be the one
that possesses the highest spectrum acquiring rate in the world. The
spectroscopic survey which contains the information about the radial
positions of galaxies, can probe the 3D distribution of galaxies
effectively. In this paper, we study the sensitivity of LAMOST to
the determination of cosmological parameters with the simulated
galaxy power spectrum. In our analysis, we also consider the
simulated observations for the future CMB and SN Ia measurements
from PLANCK and the 5-year SNLS, which are presumably conducted
during the same time period as LAMOST survey. Our results indicate
that the LAMOST has the promising potential in probing for the
cosmological parameters, especially in constraining on the EoS of
the dark energy and the neutrino mass.

The paper is organized as follows: In Section II we describe the
method of fitting and the simulation technique. In section III, we
present the results and discussions. The last section contains a
summary.

\section{Methodology}
%MCMC method

In this section, we introduce the method and the fitting procedure.
For the dynamical dark energy model, we choose the parametrization
given by \cite{Linderpara}:
\begin{equation}
\label{EoS} w(a) = w_{0} + w_{a}(1-a)~,
\end{equation}
where $a=1/(1+z)$ is the scale factor and $w_{a}=-dw/da$
characterizes the ``running" of the EoS (Run w henceforth). For the
$\Lambda$CDM model, $w_0=-1$ and $w_a=0$.

When using the MCMC global fitting strategy to constrain
cosmological parameters, dark energy perturbations should be taken
into account properly, especially for models with time evolving
EoS of dark energy. This issue has been realized by many
researchers including the WMAP group
\cite{Zhao:2005vj,globf05,wmap3,pertother}. However, when the
parameterized EoS crosses $-1$, one cannot handle the dark energy
perturbations based on quintessence, phantom, k-essence and other
non-crossing models. By virtue of quintom \cite{Quintom}, the
perturbations at the crossing points are continuous. Thus we have
proposed a technique to treat dark energy perturbations in the
whole parameter space. %For details of this method, we refer the
%readers to our previous papers \cite{Zhao:2005vj,globf05}.

In this study, we have modified the publicly available Markov Chain
Monte Carlo package CosmoMC\cite{CosmoMC} to include the dark energy
perturbations. For handling the parametrization of the EOS getting
across -1, firstly we introduce a small positive constant $\epsilon$
to divide the full range of the allowed value of the EOS $w$ into
three parts: 1) $ w
> -1 + \epsilon$; 2) $-1 + \epsilon \geq w  \geq-1 - \epsilon$; and
 3) $w < -1 -\epsilon $.
Working in the conformal Newtonian gauge, the perturbations of DE
can be described by \bea
    \dot\delta&=&-(1+w)(\theta-3\dot{\Phi})
    -3\mathcal{H}(c_{s}^2-w)\delta~~, \label{dotdelta}\\
\dot\theta&=&-\mathcal{H}(1-3w)\theta-\frac{\dot{w}}{1+w}\theta
    +k^{2}(\frac{c_{s}^2\delta}{{1+w}}+ \Psi)~~ . \label{dottheta}
\eea

Neglecting the entropy perturbation, for the regions 1) and 3), the
EOS does not across $-1$ and the perturbation is well defined by
solving Eqs.(\ref{dotdelta},\ref{dottheta}). For the case 2), the
perturbation of energy density $\delta$ and divergence of velocity,
$\theta$, and the derivatives of $\delta$ and $\theta$ are finite
and continuous for the realistic quintom DE models. However for the
perturbations of the parameterizations, there is clearly a
divergence. In our study for such a regime, we match the
perturbations in region 2) to the regions 1) and 3) at the boundary
and set
\begin{equation}\label{dotx}
  \dot{\delta}=0 ~~,~~\dot{\theta}=0 .
\end{equation}
In our numerical calculations we limit the range to be $|\Delta w =
\epsilon |<10^{-5}$ and find our method to be a very good
approximation to the multi-field quintom. More detailed treatments
can be found in Ref.\cite{Zhao:2005vj,globf05}.

Furthermore, we assume purely adiabatic initial conditions and a
flat universe.
 The parameter space we begin with for the numerical
calculation is:
\begin{equation}
\label{para1} {\bf P} \equiv \left(\omega_{b}, \omega_{c},
\Theta_{s}, \tau, w_{0}, w_{a}, n_{s}, \ln(10^{10}A_{s})
\right)~,
\end{equation}
where $\omega_{b}\equiv\Omega_{b}h^{2}$ and
$\omega_{c}\equiv\Omega_{c}h^{2}$ with $\Omega_b$ and $\Omega_c$
being the baryon and cold dark matter densities relative to the
critical density, respectively, $\Theta_{s}$ is the ratio
(multiplied by 100) of the sound horizon at decoupling to the
angular diameter distance to the last scattering surface, and $\tau$
is the optical depth.
% $f_{\nu}$ is the dark matter neutrino fraction
%at present, namely,
%\begin{equation}
%f_{\nu}\equiv\frac{\rho_{\nu}}{\rho_{DM}}=\frac{\Sigma
%m_{\nu}}{93.105~eV~\Omega_ch^2}~,
%\end{equation}
In Eq.(\ref{para1}), $A_{s}$ and $n_{s}$ characterize the power
spectrum of primordial scalar perturbations. For the pivot scale of
the primordial spectrum we set $k_{\ast} = 0.05$Mpc$^{-1}$.

In our calculations, we take the total likelihood to be the
products of the separate likelihoods (${\bf \cal{L}}_i$) of CMB,
LSS and SNIa. Defining $\chi_{L,i}^2 \equiv -2 \log {\bf
\cal{L}}_i$, we then have \be\label{chi2} \chi^2_{L,total} =
\chi^2_{L,CMB} + \chi^2_{L,LSS} + \chi^2_{L,SNIa}~. \ee If the
likelihood function is Gaussian, $\chi^2_{L}$ coincides with the
usual definition of $\chi^2$ up to an additive constant
corresponding to the logarithm of the normalization factor of
${\cal L}$.

The data used for current constraints include the three-year WMAP
(WMAP3)\footnote {With the newly released 5-year WMAP
data\cite{wmap5,komastue}, we have checked that the new data will
not significantly change the results. } Temperature-Temperature (TT)
and Temperature-Polarization (TE) power spectrum
\cite{wmap3:2006:1,wmap3:2006:2,wmap3:2006:3,wmap3:2006:4} as well
as the smaller scale experiments, including Boomerang-2K2
\cite{MacTavish:2005yk}, CBI \cite{Readhead:2004gy}, VSA
\cite{Dickinson:2004yr} and ACBAR \cite{Kuo:2002ua}, the SDSS
luminous red galaxy (LRG) sample \cite{Tegmark:2006az} and 2dFGRS
\cite{Cole:2005sx}, and recently released ESSENCE (192 sample) data
\cite{Miknaitis:2007jd,Davis:2007na}. For the LSS power spectrum, we
only use the data in the linear regime up to $k \sim 0.1 h
Mpc^{-1}$. In the calculation of the likelihood from SNIa we have
marginalized over the nuisance parameter \cite{DiPietro:2002cz}.
Furthermore, we make use of the Hubble Space Telescope (HST)
measurement of the Hubble parameter $H_{0}\equiv
100$h~km~s$^{-1}$~Mpc$^{-1}$ \cite{Hubble} by multiplying the
likelihood by a Gaussian likelihood function centered around
$h=0.72$ and with a standard deviation $\sigma=0.08$. We also impose
a weak Gaussian prior on the baryon density
$\Omega_{b}h^{2}=0.022\pm0.002$ (1 $\sigma$) from the Big Bang
Nucleosynthesis \cite{BBN}, and a cosmic age tophat prior as 10 Gyr
$< t_0 <$ 20 Gyr.

%how to generate the data
For the future data, we consider the measurements of LSS from
LAMOST, the CMB from PLANCK \cite{PLANCK} and the SN Ia from 5-year
SNLS\cite{SNLS}.

For the simulation of LAMOST, we mainly simulate the galaxy power
spectrum. We consider two sources of statical errors on the power
spectrum measurements: the sample variance and the shot noises which
due to the limited number of independent wavenumber sampled from a
finite survey volume and the imperfect sampling of fluctuations by
the finite number of the galaxies respectively,\cite{9304022} \be
\label{eqn:dPK} (\frac{\sigma_P}{P})^2 = 2\times \frac{(2
\pi)^3}{V}\times \frac{1}{4 \pi k^2 \Delta k}\times (1+
\frac{1}{\bar{n}P})^2~, \ee where $V$ is the survey volume and
$\bar{n}$ is the mean galaxy density. From the more conservative
estimation, we know that the redshift distribution of main sample of
LAMOST is between 0 and 0.6 and the mean redshift is around 0.2. So
for simplicity, in our study, we simulate the power spectrum of the
galaxies at $z=0.2$ that can be got from these galaxies. The survey
area is $15000$ $deg^2$ and the total number of galaxies within the
survey volume is $10^7$\cite{lamost}. We only consider the linear
regime, namely the maximum k we consider is $k \sim 0.1$ $h$
Mpc$^{-1}$. As we know that, the galaxy power spectrum P(k) in EQ.
(\ref{eqn:dPK}) is \be \label{bias}P(k)=b^2 p_m(k),\ee where
$p_m(k)$ is the linear matter power spectrum, and here we take $b$
as a constant $b=1$ when simulating the data and when using the
galaxy power spectrum to constrain cosmological parameters, we take
$b$ as a free parameter and marginalize over it.

For the simulation with PLANCK, we follow the method given in our
previous paper \cite{07081111}. We mock the CMB TT, EE and TE
power spectrum by assuming the certain fiducial cosmological
model. For the detailed techniques, please see our previous
companion paper \cite{07081111}. We have also simulated $500$ SN
Ia according to the forecast distribution of the SNLS
\cite{future-snls}. For the error, we follow the Ref.\cite{kim}
which takes the magnitude dispersion $0.15$ and the systematic
error $\sigma_{sys}=0.02\times z/1.7$. The whole error for each
data is given by:
\begin{equation}
\sigma_{maga}(z_i)=\sqrt{\sigma^2_{sys}(z_i)+\frac{0.15^2}{n_i}}~,\label{snap}
\end{equation}
where $n_i$ is the number of supernova of the $i'$th redshift bin.

As pointed out in our previous works
\cite{Xia:2006cr,Xia:2006wd,Zhao:2006qg}, the cosmological
parameters are highly affected by the dark energy models due to
the degeneracies among the EoS of DE and other parameters.
Therefore, in our study of this paper, we choose two fiducial
models with different dark energy properties: $\Lambda$CDM model
(fiducial model I henceforth) and dynamical dark energy (fiducial
model II henceforth) with time evolving EoS. The parameters of the
two sets of fiducial models are obtained from the current
observational data.

%\subsection{}

%\subsection{Global fitting program}

\section{Results and Discussions}
\begin{table}{\footnotesize
TABLE I. Constraints on cosmological parameters from the current
observations and the future simulations.  For the current
constraints we have shown the mean values $1\sigma$ (Mean) and the
best fit results together. For the future mocked data we list the
standard deviation (SD) of these parameters with fiducial model I (FMI)
and fiducial model II (FMII). In order to highlight
the contribution from LAMOST, we compare the results with/without
LAMOST.
\begin{center}

\begin{tabular}{|c|cccc|cccc|}

\hline

&\multicolumn{2}{c|}{Current for
$\Lambda$CDM}&\multicolumn{2}{c|}{~Future~(SD with
FMI)}&\multicolumn{2}{c|}{Current for Run
w}&\multicolumn{2}{c|}{~Future~(SD
with FMII)}\\

\cline{2-9}

&\multicolumn{1}{c|}{~Best~Fit~}&\multicolumn{1}{c|}{~~~~Mean~~~~}&\multicolumn{1}{c|}{PLANCK$+$
SNLS}&\multicolumn{1}{c|}{PLANCK$+$
SNLS}&\multicolumn{1}{c|}{~Best~Fit~}&\multicolumn{1}{c|}{~~~~Mean~~~~}
&\multicolumn{1}{c|}{PLANCK$+$SNLS}&\multicolumn{1}{c|}{PLANCK$+$
SNLS} \\

&\multicolumn{1}{c|}{}&\multicolumn{1}{c|}{}&\multicolumn{1}{c|}{}&$+$LAMOST&\multicolumn{1}{c|}{}&\multicolumn{1}{c|}{}&\multicolumn{1}{c|}{}&$+$LAMOST\\

\hline

$w_0$&$-1$&$-1$&$0.118$&$0.100$&$-1.16$&$-1.03^{+0.15}_{-0.15}$&$0.0899$&$0.0598$\\

\hline

$w_a$&$0$&$0$&$0.522$&$0.417$&$0.968$&$0.405^{+0.562}_{-0.587}$&$0.332$&$0.162$\\

\hline

$~\Omega_{\Lambda}~$&$0.760$&$0.762^{+0.015}_{-0.015}$&$0.0115$&$0.00547$&$0.756$&$0.760^{+0.017}_{-0.018}$&$0.0125$&$0.00460$\\

\hline

$H_0$&$73.1$&$73.3^{+1.6}_{-1.7}$&$1.594$&$0.828$&$70.3$&$71.2^{+2.3}_{-2.3}$&$1.840$&$0.673$\\

\hline

$\sigma_8$&$0.769$&$0.755\pm0.031$&$0.0223$&$0.0174$&$0.634$&$0.675\pm0.068$&$0.0299$&$0.0220$\\

\hline

%$\sum M_{\nu}$&$0.400$&$<0.958(95\%)$&$-$&$-$&$0.492$&$<1.59(95\%)$&$0.301$&$0.104$\\

%\hline

%$\Delta\chi^2$&\multicolumn{3}{c|}{$0$}&\multicolumn{3}{c|}{$-3.0$}&\multicolumn{2}{c|}{$-1.0$}\\

%\hline
\end{tabular}
\end{center}}
\end{table}

%In our numerical calculations, we firstly consider a simple case where
%$\alpha_s$, $r$ = 0 and also $f_{\nu}=0$ in EQ.
%(\ref{para1}).
In Table I, we present the numerical results of the constraints on
the cosmological parameters from the current data and the error
forecast from the simulated data. To show the importance of LAMOST,
we compare the two sets of results, one from PLANCK $+$ SNLS, the
other from  PLANCK $+$ SNLS $+$ LAMOST. As we know, the matter power
spectrum is directly related to the horizon size at matter-radiation
equality, in turn the matter power spectrum will make accurate
measurement of $\Omega_m h$. On the other hand, there are
degeneracies between $\Omega_m h$ and the other cosmological
parameters, {\it e.g.} $\Omega_{\Lambda}$, $H_0$, $w_0$, $w_a$ and
so on, hence the tight constraint on $\Omega_m h$ will be helpful
for breaking these degeneracies and improve the constraints on these
cosmological parameters. For example, in Table I, one can find the
constraints on $\Omega_{\Lambda}$ and $H_0$ are tightened a lot by
including LAMOST. Also from Figure 1, one can see the constraits on
the age of universe is also shrank obviously, this is because the
age is directly related to the hubble constant and $\Omega_m$.

\begin{figure}[htbp]
\begin{center}
\includegraphics[scale=0.9]{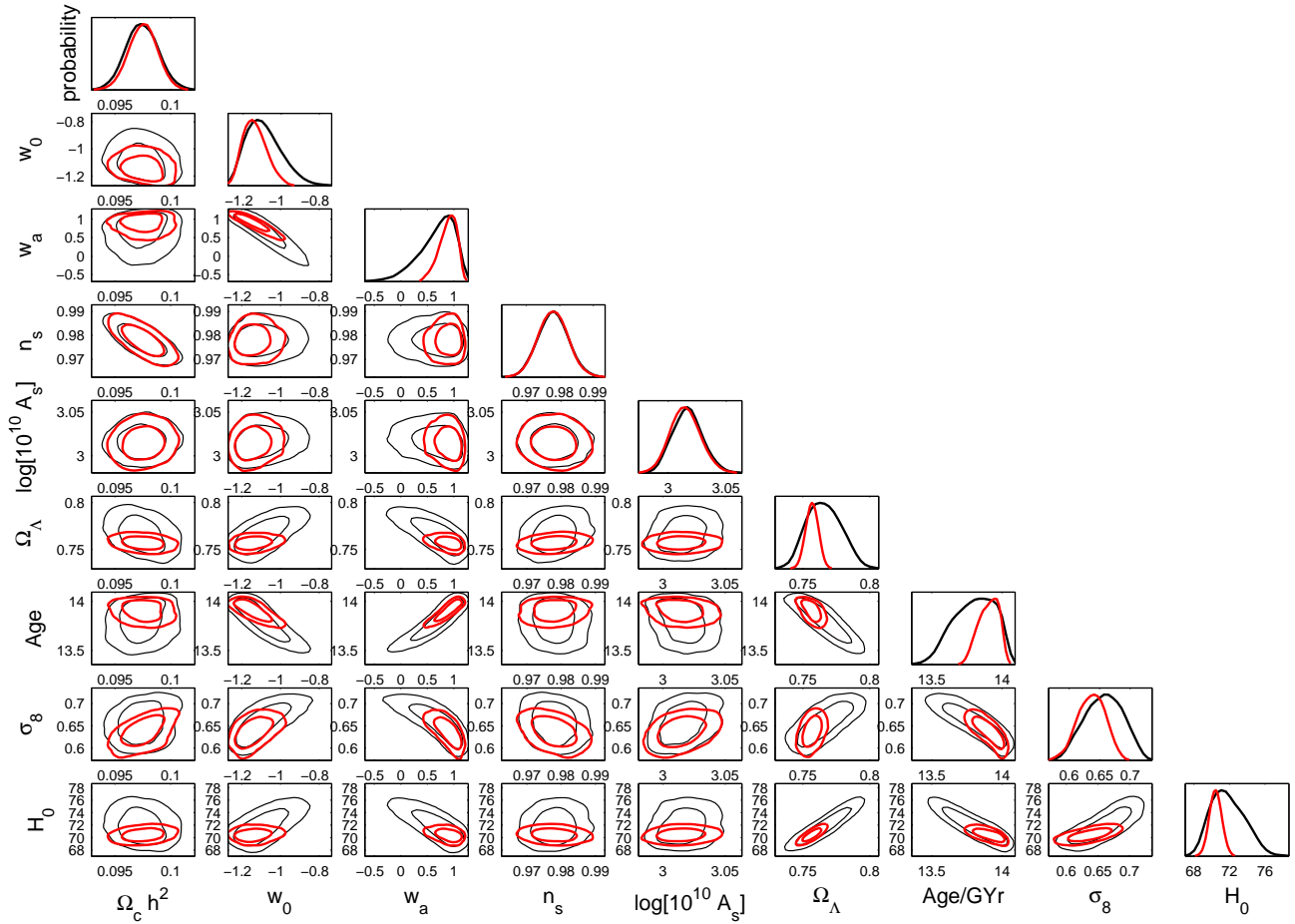}
\vskip-1.3cm \vspace{10mm}\caption{one-dimensional distributions and
two-dimensional $68\%$ and $95\%$ limits on the cosmological
parameters. The black solid lines are obtained with the simulated
PLANCK $+$ SN Ia and the red solid lines are from PLANCK $+$ SN Ia
$+$ LAMOST.\label{fig1}}
\end{center}
\end{figure}

In figure \ref{fig1}, we plot the $2-$D cross correlation and $1-$D
probability distribution of some of the basic cosmological
parameters in Eq.(\ref{para1}) and also some of the reduced
parameters. The black solid lines are given by fitting with the
simulated PLANCK and SNLS, and the red solid lines are provided by
including the simulated LAMOST data. From the comparison, we find
that LAMOST have promising potentials in constraining cosmological
parameters, such as, the EoS of dark energy, the dark energy density
budget $\Omega_{\Lambda}$, the age of universe, $\sigma_8$ and the
Hubble constant.

\begin{figure}[htbp]
\vspace{3mm}
\begin{center}
\includegraphics[scale=0.5]{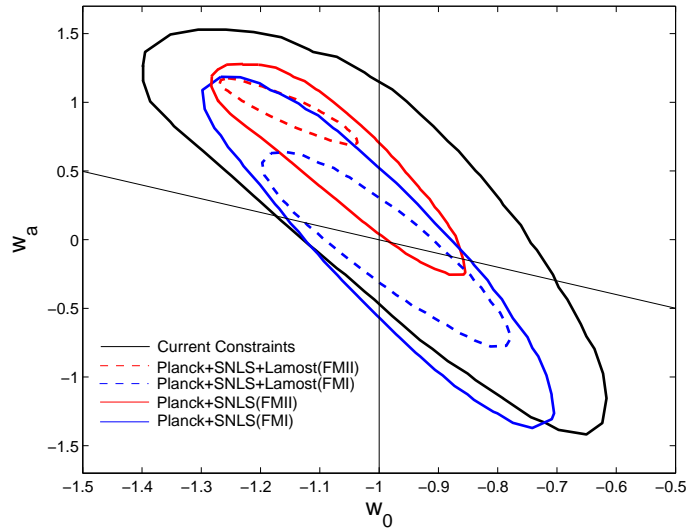}
\vskip-1.3cm \vspace{10mm}\caption{2-d joint $68\%$ and $95\%$
confidence regions of the parameters $w_0$ and $w_a$ for flat
universe. The black solid line is given by the current
constraints, the red solid line comes from the simulated data of
PLANCK and 5year SNLS with fiducial model II, while the red dashed
line is by combining the simulated LAMOST data. The blue solid
line and blue dashed line are the results with the fiducial model
I with/without LAMOST respectively.\label{fig2}}
\end{center}
\end{figure}

In order to see explicitly the effect of LAMOST on dark energy
constraints, in figure \ref{fig2}, we plot the $2\sigma$
confidence level contours on $w_0$ and $w_a$. The black solid line
is given by the current constraints and the red solid line is
given by fitting with the simulated data of PLANCK and SNLS 5 year
data with fiducial model II, while the red dashed line is given by
including the simulated LAMOST data. This comparison shows clearly
LAMOST will contribute significantly in tightening the constraints
on the EoS of dark energy. Numerically we find the best fit model
with the current data is given by the dynamical quintom model with
EoS across -1, however the cosmological constant is within
$1\sigma$ confidence level. The future PLANCK measurement and
5-year SNLS SN Ia will be able to distinguish the cosmological
constant from the dynamical model at $2\sigma$ confidence level,
while LAMOST can improve this sensitivity significantly at
$3.3\sigma$. The blue solid lines and blue dashed lines show the
comparison between the results with and without LAMOST for
fiducial model I.

On the other hand, for the parameter related to the inflation
models, such as $n_s$ and $A_s$, the constraints are mainly from
PLANCK, as pointed out in our previous paper\cite{07081111}. Adding
in LAMOST can not further tighten the constraints. We have also done
another analysis with the additional parameters $\alpha_s$ and $r$,
and obtained the similar conclusion, where $\alpha_s$ characterizes
the running of the primordial power spectrum index and $r$ is the
ratio of tensor to scalar perturbations.

Now we study the cosmological constraint on the neutrino mass by
adding in a new parameter $f_{\nu}$ in Eq.(\ref{para1}). The
parameter $f_{\nu}$ is the dark matter neutrino fraction at present,
namely,
\begin{equation}
f_{\nu}\equiv\frac{\rho_{\nu}}{\rho_{DM}}=\frac{\Sigma
m_{\nu}}{93.105~eV~\Omega_ch^2}~,
\end{equation}
where $\Sigma m_{\nu}$ is the sum of the neutrino masses. In this
study, the mocked data we use are generated by assuming the massless
neutrino {\it i.e.} $f_{\nu} = 0$ in the fiducial models.
Consequently the constraints on $f_{\nu}$ should be regarded as the
upper limits of the neutrino mass which the future observations will
be sensitive to. It is well known that the massive neutrinos modify
the shape and amplitude of the matter power spectrum, and also the
epoch of matter-radiation equality, angular diameter distance to the
last scattering surface. Thus they leave imprints on the
observations of CMB and LSS. In Table II, we provide the constraints
on neutrino mass from the current observations and the future
simulated data. For the current data\footnote{Usually Lyman-$\alpha$
data will give stringent constraint on the neutrino mass, however,
its systematics are quite unclear currently. In our global analysis
to be conservative, we have not included it \cite{silk,komastue}.},
within the framework of the $\Lambda$CDM model, we get $\Sigma
m_{\nu}<0.958~eV~(95\%)$ which is consistent with the result in
Ref.\cite{Tegmark:2006az}. For the time evolving EoS of dark energy
model, this limit is relaxed to $\Sigma m_{\nu}<1.59~eV~(95\%)$, due
to the degeneracy between the dark energy parameters and the
neutrino mass\cite{Xia:2006wd,Hannestad:2005gj}.

\begin{table}%\label{table1}
TABLE II. Constraints on neutrino mass from the current observations
and the future simulations. We have shown the $2\sigma$ upper
limits. In order to highlight the contribution from LAMOST, we
compare the results with/without LAMOST. \iffalse Note that when we
perform the analysis for the future surveys, the parameter space is
given by Eq. (\ref{para1}) plus $f_{\nu}$ for the two sets of
fiducial models.\fi
\begin{center}

\begin{tabular}{|c|c|c|c|c|}

\hline

\multicolumn{2}{|c|}{Current}&\multicolumn{3}{c|}{Future}\\

\cline{3-5}

\multicolumn{2}{|c|}{}&&PLANCK+SNLS&PLANCK+SNLS+LAMOST\\

\hline

$\Lambda$CDM&$<$0.958eV&FMI&$<$0.957eV&$<$0.377eV\\

Run w&$<$1.59eV&FMII&$<$0.915eV&$<$0.346eV \\ \hline

\hline

%$\Delta\chi^2$&\multicolumn{3}{c|}{$0$}&\multicolumn{3}{c|}{$-3.0$}&\multicolumn{2}{c|}{$-1.0$}\\

%\hline
\end{tabular}
\end{center}
\end{table}

\begin{figure}[htbp]
\begin{center}
\includegraphics[scale=0.45]{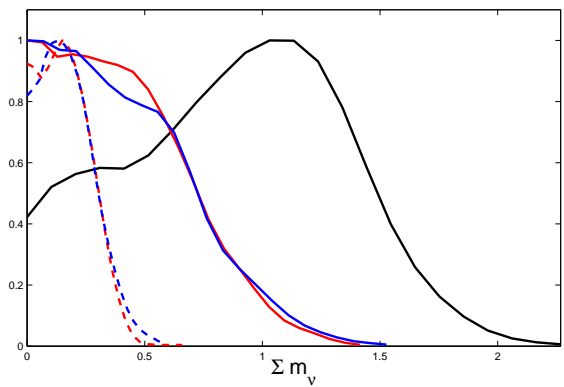}
\vskip-1.3cm \vspace{10mm}\caption{one-dimensional constraints on
the neutrino mass. The black solid line is given by the current
data, the red solid line is given by fitting with the simulated
PLANCK $+$ SNLS with fiducial model II and the red dashed line is
given by including the simulated LAMOST. The blue solid line and
blue dashed line are the results obtained with the fiducial model I.
\label{fig4}}
\end{center}
\end{figure}

With the simulated data, in figure \ref{fig4}, we illustrate the one
dimensional probability distribution of the total neutrino mass
$\sum m_{\nu}$. The black solid line is given by the current
constraints, the red solid line is given by fitting with the
simulated PLANCK $+$ SNLS with fiducial model II and the red dashed
line is given by including the simulated LAMOST. The blue solid line
and blue dashed line are the results obtained with the fiducial
model I. Our results show that the LAMOST can provide a more
stringent constraint on the neutrino mass. For example, the
$2\sigma$ neutrino mass limit is changed from $0.957 eV$ to $0.377
eV$ by including the simulated LAMOST data with the fiducial model
I.

\section{Summary}

In this paper we have studied the sensitivity of LAMOST project to
the determination of the cosmological parameters. With the simulated
$3$D matter power spectrum of LAMOST, in combination with the future
PLANCK data and 5-year SNLS data, we have obtained the constraints
on the various parameters by employing the MCMC method. Our results
show the potential for LAMOST in constraining the cosmological
parameters, especially on the EoS of dark energy and the neutrino
mass.

We have performed our analysis in flat universe, however, if we take
the curvature $\Omega_k$ into consideration, namely if we free
$\Omega_k$ in the global analysis, the basic conclusion will not
change. That is to say, we can also find the potential of the future
LAMOST data in determining cosmological parameters, however, the
specific contours of each cosmological parameters will be enlarged
for the additional degree of freedom, and more relevant discussion
can be seen in our previous paper\cite{Zhao:2006qg}, in which we
have implemented the global fitting with the observational data for
non-flat universe.
%In our
%study, we use two fiducial model in simulating the future data and
%provide the further constraints on the cosmological parameters
%respectively, we find the final results are weakly dependent on
%the choice of the fiducial model.

{\it Acknowledgement. --- }

Our MCMC chains were finished in the Shuguang 4000A of the Shanghai
Supercomputer Center (SSC). We thank Xuelei Chen, Bo Qin, Charling
Tao, Lifan Wang, Pengjie Zhang, Yong-Heng Zhao, Zong-Hong Zhu,
Tao-Tao Qiu and Lei Sun for discussions. This work is supported in
part by China postdoctoral science foundation, National Science
Foundation of China under Grant No. 10533010, and the 973 program
No.2007CB815401, and by the Key Grant Project of Chinese Ministry of
Education (No. 305001).

\end{document}